\documentclass[reprint,amsmath,amssymb,aps,]{revtex4-2}

\usepackage{graphicx}
\usepackage{dcolumn}
\usepackage{bm}

\usepackage[]{hyperref}
\usepackage{color} 
\usepackage{amsmath}

\begin{document}

\title{High-efficiency telecom frequency conversion via a diamond-type atomic ensemble}

\author{Ling-Chun Chen,$^{1,2}$ Meng-Yi Lin,$^{1}$ Jiun-Shiuan Shiu,$^{1,2}$ Xuan-Qing Zhong,$^{1}$ Po-Han Tseng,$^{1}$ and Yong-Fan Chen$^{1,2,3}$}

\email{yfchen@mail.ncku.edu.tw}

\affiliation{
$^1$Department of Physics, National Cheng Kung University, Tainan 70101, Taiwan\\ 
$^2$Center for Quantum Frontiers of Research $\&$ Technology, Tainan 70101, Taiwan\\
$^3$Center for Quantum Science and Technology, National Tsing Hua University, Hsinchu, Taiwan
}



\begin{abstract}

Efficient telecom frequency conversion (TFC) in atomic systems is crucial for integrating atom-based quantum nodes into low-loss fiber-optic quantum networks. Here, we demonstrate high-efficiency TFC from 795 nm to 1367 nm in a cold $^{87}$Rb ensemble via diamond-type four-wave mixing (FWM), achieving conversion efficiencies of 66\% and 80\% at optical depths of 75 and 110, respectively, using a weak coherent probe field. These results surpass all previously reported values in atomic systems, enabled by a systematic investigation of the built-in V-type and cascade-type electromagnetically induced transparency spectra that guided the optimization of FWM conditions. Although this work employs coherent fields, our previous theoretical study has shown that quantum states can be preserved with high fidelity during the conversion process, highlighting the promise of diamond-type atomic FWM as a robust interface for long-distance quantum communication.

\end{abstract}


\maketitle


\newcommand{\FigOne}{
    \begin{figure}[t]
    \centering
    \includegraphics[width = 8.8 cm]{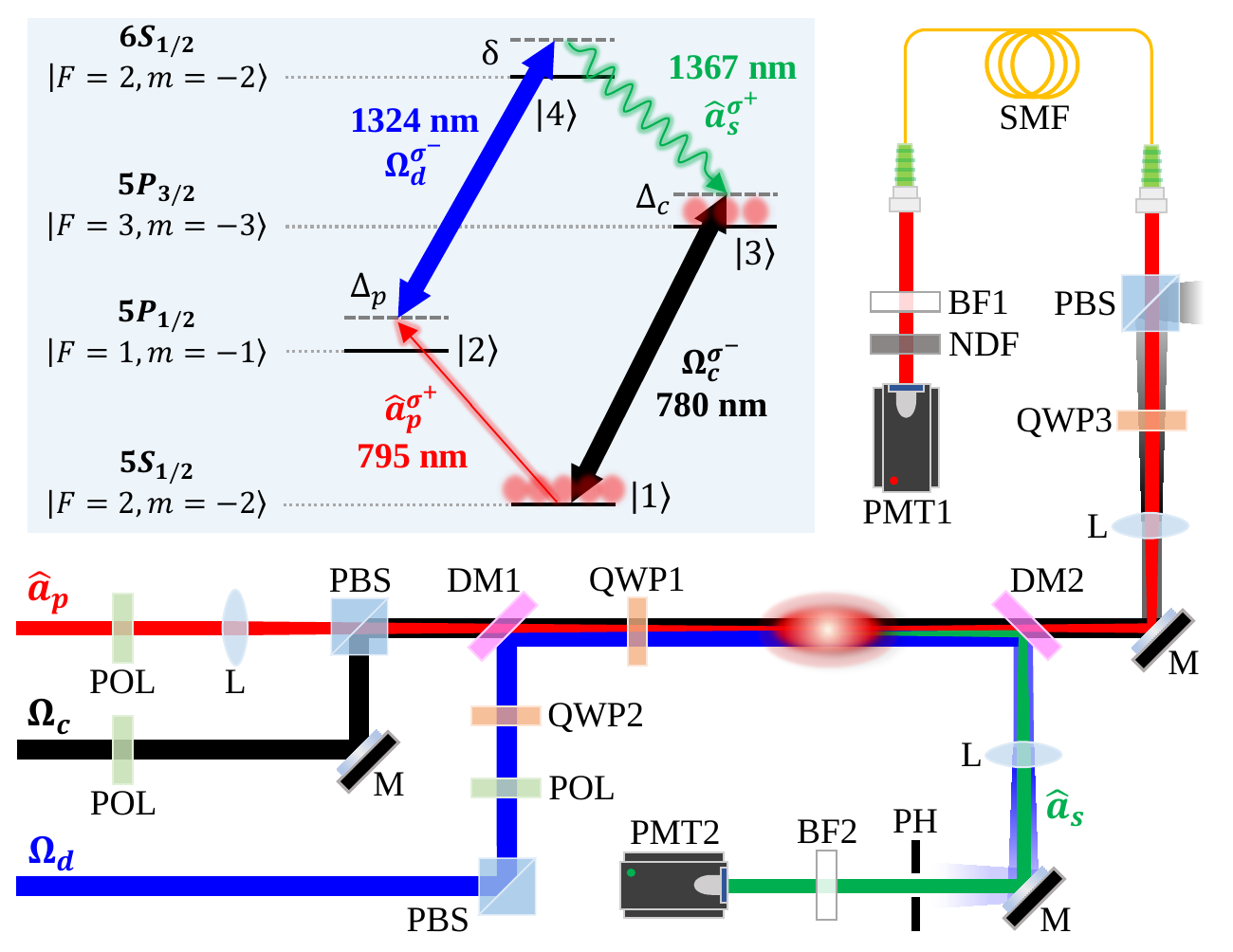}
    \caption{
Diagram of the diamond-type FWM and experimental setup. $\hat{a}^{\sigma^+}_{p}$, $\Omega^{\sigma^-}_{c}$, $\Omega^{\sigma^-}_{d}$ and $\hat{a}^{\sigma^+}_{s}$ represent probe, coupling, driving and signal fields. POL, polarizer; M, mirror; L, lens; PBS, polarizing beam splitter; QWP, quarter-wave plate; DM, dichroic mirror; PH, pinhole; BF, bandpass filter; NDF, neutral density filter; SMF, single-mode fiber; PMT, photomultiplier tube.
}
    \label{fig:setup}
    \end{figure}
}

\newcommand{\FigTwo}{
    \begin{figure}[t]
    \centering
    \includegraphics[width = 8.6 cm]{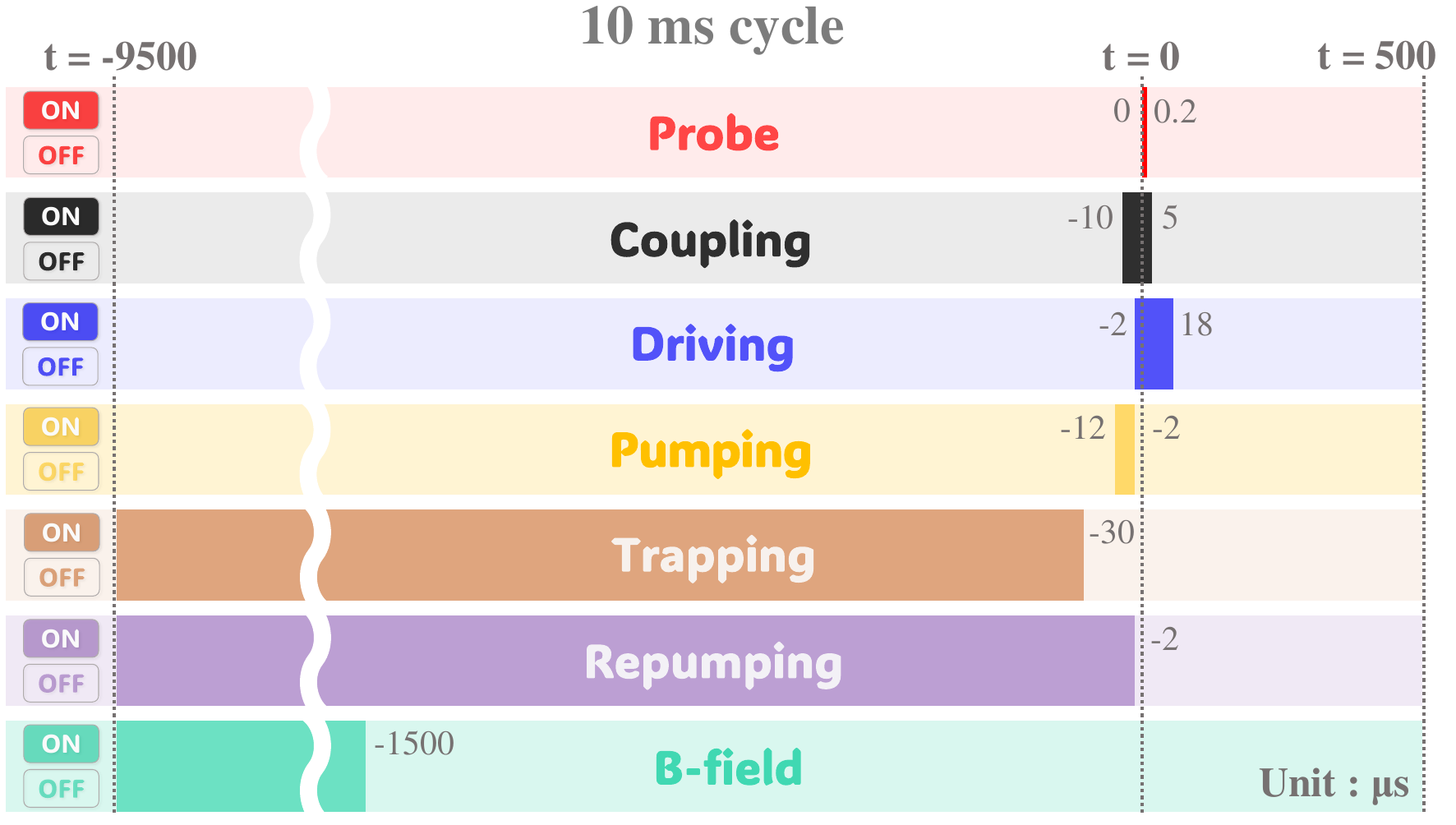}
    \caption{
Timing sequence for the diamond-type FWM experiment. Each cycle spans 10\,ms. The time origin ($t=0$) marks the onset of the probe pulse.
}
    \label{fig:timing sequence}
    \end{figure}
}

\newcommand{\FigThree}{
    \begin{figure}[t]
    \centering
    \includegraphics[width = 9.0 cm]{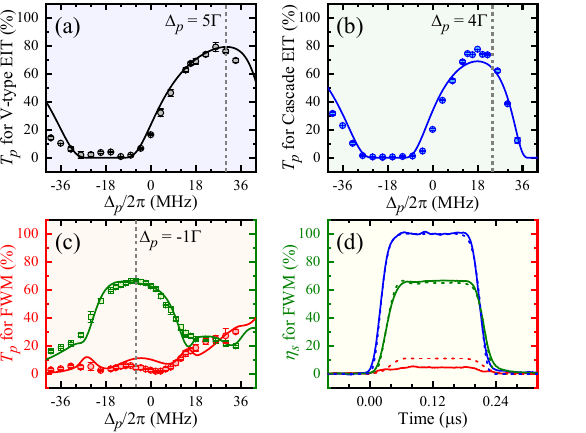}
    \caption{
Spectra in the bright MOT with $\mathrm{OD}=75$, $\Omega_{c}=11\Gamma$, $\Omega_{d}=9\Gamma$, $\Delta_{c}=5\Gamma$, and $\Delta_{d}=-4\Gamma$. Circles denote experimental data and curves are theoretical predictions. Black, blue, and red lines show the probe transmission $T_p$ for (a) V-type EIT, (b) cascade-type EIT, and (c) diamond-type FWM, respectively, while green lines and squares represent the predicted and measured signal conversion efficiency $\eta_s$. (d) shows the time-domain pulse profiles corresponding to the peak conversion efficiency in (c), at $\Delta_{p} = -1\Gamma$, where $\eta_s \approx 66\%$. The traces include the incident probe (blue), transmitted probe (red), and generated signal (green). All data are normalized, and the vertical axis represents the probe transmission and signal conversion efficiency. Dashed lines indicate the corresponding theoretical curves.
}
    \label{fig:spectra BM}
    \end{figure}
}

\newcommand{\FigFour}{
    \begin{figure}[t]
    \centering
    \includegraphics[width = 9.0 cm]{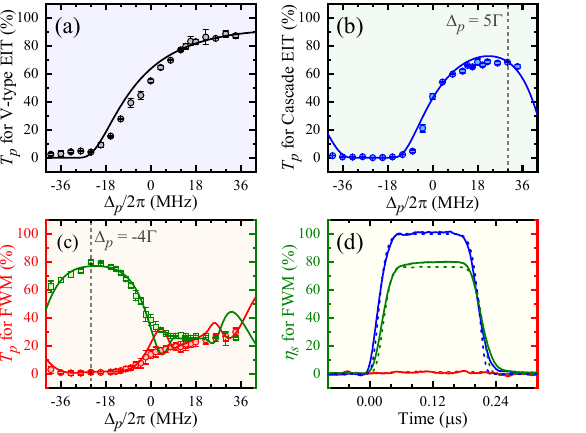}
    \caption{
Spectra in the dark SPOT with $\mathrm{OD}=110$, $\Omega_{c}=20\Gamma$, $\Omega_{d}=12\Gamma$, $\Delta_{c}=8\Gamma$, and $\Delta_{d}=-5\Gamma$. Circles denote experimental data and curves are theoretical predictions. Black, blue, and red lines show the probe transmission $T_p$ for (a) V-type EIT, (b) cascade-type EIT, and (c) diamond-type FWM, respectively, while green lines and squares represent the predicted and measured signal conversion efficiency $\eta_s$. (d) shows the time-domain pulse profiles corresponding to the peak conversion efficiency in (c), at $\Delta_{p} = -4\Gamma$, where $\eta_s \approx 80\%$. The traces include the incident probe (blue), transmitted probe (red), and generated signal (green). All data are normalized, and the vertical axis represents the probe transmission and signal conversion efficiency. Dashed lines indicate the corresponding theoretical curves.
}
    \label{fig:spectra DM}
    \end{figure}
}


\section{INTRODUCTION} \label{sec:Introduction}

Quantum technologies, including quantum networks~\cite{network1,network2,network3,network4}, quantum cryptography~\cite{cryp1,cryp2,cryp3}, and quantum teleportation~\cite{tele1,tele2,tele3,tele4}, have become fundamental to modern advancements. These protocols can be efficiently implemented using photons, which serve as ideal carriers of quantum information due to their weak interaction with the environment, rapid transmission, and highly tunable properties such as polarization and frequency. Among these properties, telecom wavelengths—particularly 1550 nm (C-band) and 1310 nm (O-band)—are crucial for minimizing photon attenuation in optical fibers, as they offer significantly lower losses than visible wavelengths and are thus optimal for long-distance fiber-based quantum communication. Consequently, telecom-wavelength single-photon sources have been extensively explored~\cite{review}. However, integrating telecom-band photons with atom-based quantum devices remains challenging, as the transition frequencies of alkali atoms, such as the rubidium D$_1$ (795 nm) and D$_2$ (780 nm) lines, do not naturally match telecom wavelengths. Therefore, developing high-fidelity telecom frequency conversion (TFC)~\cite{Qubittransfer,Jen,PoHan} is essential for bridging disparate photon wavelengths and enabling efficient interfacing between atom-based quantum systems and long-distance quantum communication over optical fibers.

As photonic quantum technologies continue to advance, TFC has become an important technique for enhancing the flexibility and performance of scalable quantum networks. For example, the Duan–Lukin–Cirac–Zoller (DLCZ) protocol~\cite{DLCZ}, which underpins a robust quantum repeater architecture, can benefit from TFC when transmitting write photons to intermediary stations for Bell-state measurements. While the DLCZ protocol can be implemented using a dense repeater architecture, TFC provides an alternative means of mitigating fiber loss by enabling low-loss transmission in the telecom band. In fact, single-photon sources operating in the near-infrared regime can be shifted to the telecom band via quantum frequency conversion~\cite{QFC} while preserving high fidelity~\cite{TQFC1,TQFC2,TQFC3}. Moreover, TFC plays a crucial role in applications by expanding the degrees of freedom of photon sources. This includes the development of quantum networks based on entangled photon pairs~\cite{DLCZ Realization} and the realization of a photonic quantum interface using orbital angular momentum~\cite{OAM QFC}.

Over the years, a variety of TFC schemes have been explored. Atomic systems have been implemented in both warm vapor cells~\cite{warm1,warm2,warm3} and cold atomic ensembles~\cite{CE54,CE32,CE65}, while solid-state platforms—particularly periodically poled lithium niobate (PPLN) waveguides—have also been widely employed~\cite{PPLN1,PPLN2,PPLN3,PPLN4}. More recently, molecular hydrogen has emerged as an alternative medium for TFC, with implementations including pressurized hydrogen cells~\cite{hydrogen cell} and hydrogen-filled hollow-core fibers~\cite{HFHCF}. Additionally, hollow-core photonic crystal fibers (HCPCFs) filled with rubidium vapor have shown promise for efficient telecom conversion~\cite{HCPCFs}. Among these approaches, the highest conversion efficiency reported in cold atomic ensembles is 54\% based on rigorous experimental verification~\cite{CE54}, whereas a subsequent study reported that under copolarized conditions efficiencies of up to 65\% were observed, albeit without detailed experimental data~\cite{CE65}. In contrast, nonlinear conversion in PPLN, benefiting from strong nonlinear susceptibility and tight mode confinement, has reached efficiencies as high as 96\%~\cite{PPLN3}.

Although the high efficiencies achieved in PPLN are impressive, atomic systems remain attractive for interfacing with quantum memories and other atom-based devices. Our previous theoretical work investigated ensemble-based diamond-type four-wave mixing (FWM) for TFC, demonstrating its potential to preserve photonic quantum states~\cite{PoHan}. Theoretical predictions indicate that the signal conversion efficiency $\eta_s$ can exceed 90\% under conditions of high optical depth (OD) and strong applied optical fields. Under more practical conditions, efficiencies above 80\% are achievable; however, these remain highly sensitive to parameters such as OD, optical power, and detunings. Consequently, the highest reported telecom-band conversion efficiency in atomic systems remains at 65\%, achieved with an OD of approximately 150, yet it still falls short of the theoretical potential~\cite{PoHan}.

In this study, we demonstrate high-efficiency TFC using a weak coherent probe field via diamond-type FWM in a $^{87}$Rb atomic ensemble. By optimizing operational conditions, we achieve signal conversion efficiencies $\eta_s$ of 66\% and 80\% at OD values of 75 and 110, respectively, thereby surpassing previously reported results. We further characterize the spectra of two built-in electromagnetically induced transparency (EIT) mechanisms—namely, the V-type~\cite{V1,V2} and cascade-type~\cite{cascade1,cascade2}—and provide a clear physical explanation for the observed behavior. These findings not only enhance TFC efficiency but also establish a robust foundation for the development of highly efficient quantum optical technologies, thereby advancing the prospects of quantum communication and networking.

\FigOne


\section{EXPERIMENTAL SETUP} \label{sec:Setup}

A cigar-shaped ensemble of $^{87}$Rb atoms is prepared in a magneto-optical trap (MOT) using red-detuned (by 25\,MHz) trapping and repumping lasers. The trapping laser addresses the $|5S_{1/2}, F=2\rangle \leftrightarrow |5P_{3/2}, F=3\rangle$ transition for laser cooling, while the repumping laser drives the $|5S_{1/2}, F=1\rangle \leftrightarrow |5P_{3/2}, F=2\rangle$ transition to pump atoms into the $|5S_{1/2}, F=2\rangle$ ground state. In this “bright MOT” configuration, the OD at 795\,nm reaches about 75 (the length of the atomic cloud is approximately 8\,mm). To further increase OD, we implement a “dark SPOT” (dark spontaneous-force optical trap)~\cite{darkSPOT}, where the repumping beam is spatially masked at the trap center, allowing more atoms to accumulate in the dark state $|5S_{1/2}, F=1\rangle$ (the length of the atomic cloud is approximately 4\,mm). Before the FWM experiment, an additional pumping field transfers atoms to the $|5S_{1/2}, F=2\rangle$ state, raising the OD to 110 while minimizing heating and unwanted pumping effects.

A schematic diagram of the experimental setup is shown in Fig.~\ref{fig:setup}. In the diamond-type FWM experiment, three laser beams—probe, coupling, and driving—are required to generate the signal beam. The 795\,nm probe field, denoted by the quantum operator $\hat{a}_p$, and the 780\,nm coupling field, with Rabi frequency $\Omega_c$, are generated using independent external-cavity diode lasers (Toptica DL 100). A tapered amplifier (Toptica TA 100) is used to increase the optical power of $\Omega_c$, while the 1324\,nm driving field (Rabi frequency $\Omega_d$) is directly generated by an amplified laser system (Toptica TA pro). All three beams are Gaussian. The probe beam is focused to a waist of approximately 0.125\,mm, with a Rayleigh length of 62\,mm, while both the coupling and driving beams are collimated to a waist diameter of approximately 1.3\,mm, providing a larger overlap region and uniform intensity profiles. A polarizing beam splitter (PBS) combines the probe and coupling beams, and a dichroic mirror (DM1) overlaps them with the driving beam. A near-infrared quarter-wave plate (QWP1) sets the probe and coupling beams to $\sigma^+$ and $\sigma^-$ polarizations, respectively, while a telecom quarter-wave plate (QWP2) sets the driving beam to $\sigma^-$ polarization.

Under the phase-matching condition $\Delta\vec{k} = \vec{k}_p + \vec{k}_d - \vec{k}_c - \vec{k}_s = 0$, the 1367\,nm signal field ($\hat{a}_s$) propagates collinearly with the other input beams. After the interaction region, the probe and signal fields are detected by photomultiplier tubes (PMT1 and PMT2, with calibrated gains of 164\,mV/nW and 0.065\,mV/nW, respectively). During probe detection, the coupling beam is suppressed by approximately 80\,dB by allowing the probe to passthrough a dedicated bandpass filter (BF1) and by ensuring that the coupling beam diverges before reaching PMT1. The main contributors to this attenuation include the PBS (18\,dB), single-mode fiber (SMF, 14\,dB), BF1 (26\,dB), and a neutral density filter (NDF, 20\,dB). The overall detection efficiency of the probe field is approximately 0.2\%, determined primarily by the transmission rates of the PBS (78\%), SMF (60\%), BF1 (40\%), and NDF (1\%). Similarly, the driving beam is attenuated by about 60\,dB during signal detection, ensuring minimal background noise and reliable measurements. The primary attenuation mechanisms in this case include the pinhole (PH, 2\,dB) and BF2 (53\,dB), with an additional 4\,dB loss due to beam clipping caused by magnification before reaching PMT2. The detection efficiency of the signal field is approximately 75\%, primarily limited by optical transmission through the involved components.

The inset of Fig.~\ref{fig:setup} illustrates the $^{87}$Rb energy-level scheme for the diamond-type FWM process that underpins efficient TFC in this system. Coherence between states $|3\rangle$ and $|4\rangle$ is established via a cycling transition between $|1\rangle$ and $|3\rangle$, achieved by selecting specific Zeeman sublevels to eliminate population loss and maintain atomic coherence~\cite{PoHan}. The probe detuning is defined as $\Delta_p = \omega_p - \omega_{21}$, the coupling detuning as $\Delta_c = \omega_c - \omega_{31}$, the driving detuning as $\Delta_d = \omega_d - \omega_{42}$, and the two-photon detuning as $\delta = \Delta_p + \Delta_d$, where $\omega_l$ and $\omega_{jk}$ denote the laser and atomic transition frequencies, respectively. Ultimately, the 795\,nm probe photon is down-converted into a 1367\,nm signal photon through the diamond-type FWM process, thereby achieving efficient TFC.

\FigTwo

Figure~\ref{fig:timing sequence} illustrates the timing sequence of the experiment, which operates at a repetition rate of 100\,Hz with each cycle lasting 10\,ms. A 200\,ns probe pulse ensures that the broadband diamond-type FWM process reaches a steady state. To prepare the atomic population, the coupling field is activated 10\,$\mu$s prior to the probe pulse. Moreover, the magnetic field (B-field) is switched off 1.5\,ms before the probe pulse to suppress residual magnetic fields induced by eddy currents in the MOT coil. This timing sequence is maintained in both the bright MOT and dark SPOT configurations, with the latter including an additional 10\,$\mu$s pumping light pulse to transfer atoms to the desired state and optimize population preparation. Because only one 200\,ns probe pulse is applied per 10\,ms cycle, the duty cycle of the probe field is merely 0.002\%. By contrast, the cold atoms remain cooled for 9.47\,ms in each cycle, corresponding to a 94.7\% duty cycle of the trapping field. This configuration helps maintain stable atomic conditions and ensures a high OD for the FWM process.


\section{THEORETICAL MODEL}  \label{Theoretical}

In the diamond-type FWM process, the atomic ensemble state evolves according to the Heisenberg--Langevin equations (HLEs):
\begin{align}
	\frac{\partial}{\partial t}\hat{\sigma}_{jk} = \frac{i}{\hbar}[\hat{H}_{S}, \hat{\sigma}_{jk}] + \hat{R}_{jk} + \hat{F}_{jk}, \label{eq0}
\end{align}
where $\hat{\sigma}_{jk}$ represents the collective atomic operator corresponding to the $|k\rangle \rightarrow |j\rangle$ transition, $\hat{H}_{S}$ is the system Hamiltonian, $\hat{R}_{jk}$ describes the relaxation processes, and $\hat{F}_{jk}$ accounts for Langevin noise. These equations govern the atom-field interactions, incorporating both coherent evolution and noise effects. The HLEs are coupled to the Maxwell--Schrödinger equations (MSEs), which describe the propagation of light fields under the influence of the atomic ensemble. For a weak input field $\hat{a}_p$, the driving field $\Omega_d$ is treated as a constant. The dynamics of $\Omega_c$, $\hat{a}_p$, and $\hat{a}_s$, governed by their respective MSEs, describe the interaction between the atomic ensemble and the optical fields and are given by:
\begin{align}
&\left(\frac{1}{c}\frac{\partial}{\partial t}+\frac{\partial}{\partial z}\right)\Omega_c
=\frac{i\gamma_{31}\alpha_c}{2L}\langle\hat{\sigma}_{13}\rangle,
\label{eq1}
\\
&\left(\frac{1}{c}\frac{\partial}{\partial t}+\frac{\partial}{\partial z}\right)\hat{a}_p
=\frac{i\gamma_{21}\alpha_p}{4Lg_p}\hat{\sigma}_{12},
\label{eq2}
\\
&\left(\frac{1}{c}\frac{\partial}{\partial t}+\frac{\partial}{\partial z}\right)\hat{a}_s
=\frac{i\gamma_{43}\alpha_s}{4Lg_s}\hat{\sigma}_{34},
\label{eq3}
\end{align}
where $\alpha_l = n\frac{3\lambda_l^2}{2\pi}\frac{\Gamma_{jk}}{\gamma_{jk}}L$ ($l = c, p, s$) represents OD, $n$ the atomic density, $\lambda_l$ the wavelength, $\Gamma_{jk}$ the spontaneous decay rate from the state $|j\rangle$ to $|k\rangle$, $\gamma_{jk}$ the decoherence rate, $L$ the medium length, and $g_l$ the coupling strength between photons and the ensemble. Using the Fourier transform, the frequency-domain field operators $\widetilde{a}_l$ are obtained as follows:
\begin{align}
\begin{bmatrix}
\widetilde{a}_p(L,\omega) \\ \widetilde{a}_s(L,\omega)
\end{bmatrix}
&=
\begin{bmatrix}
A(\omega) & B(\omega) \\ C(\omega) & D(\omega)
\end{bmatrix}\begin{bmatrix}
\widetilde{a}_p(0,\omega) \\ \widetilde{a}_s(0,\omega)
\end{bmatrix}
\nonumber\\&+
\sqrt{\frac{N}{c}}\sum_{jk}\int_0^L
\begin{bmatrix}
P_{jk}(z,\omega) \\ Q_{jk}(z,\omega)
\end{bmatrix}
\widetilde{F}_{jk}(z,\omega)dz.
\label{eq4}
\end{align}
Here, $N$ represents the number of atoms, while $\widetilde{a}_l(z,\omega)$ and $\widetilde{F}_{jk}(z,\omega)$ denote the frequency-domain field and noise operators, respectively. The coefficients $P_{jk}$ and $Q_{jk}$ characterize the influence of the vacuum field on the FWM system~\cite{ChinYao}. Additionally, $A$ and $D$ preserve the mode, while $B$ and $C$ facilitate mode conversion~\cite{ZiYu}. At steady state ($\omega = 0$), the transmission $T_p$ of $\hat{a}_p$ and the conversion efficiency $\eta_s$ of $\hat{a}_s$ are given by:
\begin{align}
T_p=&|A(0)|^2,\label{eq5}
\\
\eta_s=&|C(0)|^2.\label{eq6}
\end{align}
Since $T_p$ and $\eta_s$ are independent of $P_{jk}$ and $Q_{jk}$, the diamond-type FWM inherently prevents quantum states from being influenced by vacuum fluctuations. When $\eta_s$ reaches unity, this FWM process perfectly preserves photonic quantum states. These features establish the scheme as a robust solution for TFC. For further details, please refer to our previously established theoretical framework~\cite{PoHan}.

\FigThree


\section{RESULTS AND DISCUSSION} \label{results}

\subsection{Optimal TFC efficiency}

Achieving high conversion efficiency in diamond-type FWM hinges on establishing strong atomic coherence between states \( |3\rangle \) and \( |4\rangle \). This coherence is generated through two key mechanisms. First, a cascade-type EIT is realized by driving the \( |1\rangle \leftrightarrow |2\rangle \) transition with a weak probe field \( \hat{a}_p \) and simultaneously applying a strong coupling field \( \Omega_d \) on the \( |2\rangle \leftrightarrow |4\rangle \) transition, thereby linking states \( |1\rangle \) and \( |4\rangle \). Second, a detuned control field \( \Omega_c \) couples states \( |1\rangle \) and \( |3\rangle \) without inducing significant depletion. The combined effect of these processes enables efficient FWM operation. Notably, the detuned \( \Omega_c \) also induces Autler-Townes splitting (ATS), which shifts the original energy levels and causes the optimal operating parameters to deviate from full resonance.

In the bright MOT configuration, the probe field’s OD, \(\alpha_{p}\), is maintained at 75. To ensure operation in the weak-field regime, the peak power of the square probe pulse is fixed at 200 nW throughout the experiment. Under these conditions, theoretical analysis predicts that the maximum conversion efficiency is achieved at \(\Omega_c = 11\Gamma\), \(\Omega_d = 9\Gamma\), \(\Delta_c = 5\Gamma\), \(\Delta_d = -4\Gamma\), and \(\Delta_p = -1\Gamma\), where \(\Gamma = 2\pi \times 6\,\mathrm{MHz}\) denotes the spontaneous decay rate of the excited state \(|3\rangle\). According to the ATS effect, the blue-detuned \(\Omega_c\) shifts the effective energy level of \(|1\rangle\) upward, resulting in a theoretically predicted optimal two-photon detuning \(\delta = \Delta_p + \Delta_d\) that is negative. Notably, the model incorporates the spatial attenuation of \(\Omega_c\), which precludes accurately estimating the magnitude of this frequency shift using a simple approximation.

In Fig.~\ref{fig:spectra BM}(a), we examine the atomic response when only the probe and coupling fields are applied (i.e., with the driving field omitted). Here, the coupling field’s optical power is set to 5.7\,mW (Rabi frequency is $11\Gamma$), corresponding to a V-type EIT setup. We selected experimental parameters that align with the optimal conditions for diamond-type FWM. The probe transmission, \(T_p\), reaches its maximum near \(\Delta_p = \Delta_c = 5\Gamma\), primarily due to the two-photon resonance between states \(|2\rangle\) and \(|3\rangle\) driven by the probe and coupling fields. This resonance establishes an alternative pathway for atoms in \(|2\rangle\) to transition to \(|3\rangle\) via a stimulated two-photon process, thereby reducing spontaneous emission and preserving the probe field. As \(\Omega_c\) increases, this pathway becomes increasingly dominant over spontaneous decay, leading to enhanced \(T_p\). Notably, the error bars shown for all experimental data represent the standard deviation calculated from 8 data points, with each point being the average of 4096 probe pulses.

Figure~\ref{fig:spectra BM}(b) presents the cascade-type EIT spectrum obtained by applying only the probe and driving fields. To maintain a constant OD, a coupling pulse is still employed but is switched off 2\,$\mu$s before the probe pulse. Under these conditions, the driving field is set to 6.3\,mW (Rabi frequency is $9\Gamma$) with a detuning of \(-4\Gamma\). Unlike in V-type EIT, the maximum probe transmission $T_p$ does not occur at $\delta = 0$ (or $\Delta_p = 4\Gamma$). This occurs because, in the cascade configuration, one-photon absorption (at $\Delta_{p} = 0$) and two-photon absorption (at $\Delta_{p} = -\Delta_{d}$) coexist and interfere, resulting in a shift of the transparency peak. This shift lies within the range $\delta = 0$ to $\delta = \Delta_{d}/2$ (or $\Delta_p = 4\Gamma$ to $2\Gamma$), depending on the relative contributions and spectral widths of the absorption features. As $\Omega_{d}$ increases, ATS broadens the transparency window and makes the interference more symmetric. Consequently, the EIT peak shifts progressively toward $\Delta_{p} = -\Delta_{d}/2$.

Figure~\ref{fig:spectra BM}(c) shows the diamond-type FWM spectrum, which displays both the probe transmission $T_p$ and the conversion efficiency $\eta_s$. Theoretical predictions exhibit pronounced oscillations in the probe spectrum, whereas the experimental results are noticeably smoother. This discrepancy is likely due to the omission of laser linewidth effects in the theoretical model, which will be discussed in the next section. In contrast, the signal spectrum is broader, making it less sensitive to the linewidth-induced smoothing. At $\Delta_p = -1\Gamma$, we achieve a conversion efficiency of $\eta_s = 66\%$, surpassing the highest reported efficiency for ensemble-based TFC. Figure~\ref{fig:spectra BM}(d) presents the corresponding probe and signal pulses under these parameters. The pulse profile confirms that the signal field reaches a steady state within 200\,ns. Notably, our previous theoretical work~\cite{PoHan} indicates that reversing the detuning signs yields the same $T_p$ and $\eta_s$ under the conditions $\Omega_c = 11\Gamma$, $\Omega_d = 9\Gamma$, $\Delta_p = 1\Gamma$, $\Delta_c = -5\Gamma$, and $\delta = 5\Gamma$; these parameters can be interpreted in light of the mechanisms discussed above. Moreover, Fig.~\ref{fig:spectra BM}(d) shows that the total photon number of the output fields $\hat{a}_p$ and $\hat{a}_s$ is lower than that of the input $\hat{a}_p$ field, primarily due to energy loss caused by spontaneous emission. The magnitude of this loss depends on the effectiveness of the $\Omega_c$ and $\Omega_d$ fields in suppressing such emission.

Interestingly, in the case of frequency up-conversion, the optimized set of five parameters and the corresponding conversion efficiencies closely resemble those of the down-conversion scenario. This finding suggests that the same parameter set can be used to maximize the conversion efficiency for both frequency down- and up-conversion processes in the diamond-type TFC system.

\FigFour


\subsection{Higher OD Conditions} 

Our theoretical model indicates that achieving high conversion efficiency $\eta_s$ requires a large OD. Accordingly, we employ a dark SPOT configuration to increase the OD to 110, as shown in Fig.~\ref{fig:spectra DM}. Under these conditions, the optimized parameters are $\Omega_{c} = 20\Gamma$, $\Omega_{d} = 12\Gamma$, $\Delta_{c} = 8\Gamma$, $\Delta_{d} = -5\Gamma$, and $\Delta_{p} = -4\Gamma$. Figures~\ref{fig:spectra DM}(a) and \ref{fig:spectra DM}(b) illustrate the corresponding V-type and cascade-type EIT spectra. In Fig.~\ref{fig:spectra DM}(c), we observe an elevated $\eta_s$ of approximately 80\% at $\Delta_{p} = -4\Gamma$. Figure ~\ref{fig:spectra DM}(d) shows the temporal profiles of the probe and signal pulses under these conditions. In this case, stronger $\Omega_{c}$ and $\Omega_{d}$ fields effectively suppress spontaneous emission, leading to reduced energy dissipation and improved preservation of the output $\hat{a}_p$ and $\hat{a}_s$ fields. Notably, according to theoretical predictions, for $\Delta_{p} > 0$, both $T_p$ and $\eta_s$ exhibit oscillatory structures with frequencies below 10\,MHz. These features become more pronounced at high OD, where the spectral response is more sensitive to detunings. However, these oscillatory features are not observed in the experimental data. We speculate that this effect may arise from the combined laser linewidth of $\hat{a}_p$, $\Omega_c$, and $\Omega_d$ (approximately 5 MHz), which causes spectral broadening that can smear out fine oscillatory features. In addition, inhomogeneities in the cold atomic medium and possible deviations from ideal Gaussian beam profiles may further attenuate these oscillations, rendering them less discernible. Nevertheless, the overall spectral trends remain consistent with the theoretical predictions, and the agreement between experiment and theory in both V-type and cascade-type EIT spectra corroborates these observations.

To further improve $\eta_s$ at a fixed OD, one can increase $\Omega_c$ and $\Omega_d$ while carefully tuning the detunings. However, not all combinations of Rabi frequencies and detunings yield high $\eta_s$, as $\hat{a}_s$ can revert to $\hat{a}_p$ via FWM interactions during propagation. This underscores the importance of selecting distinct parameters in the bright MOT and dark SPOT configurations for optimizing $\eta_s$. 

Beyond the down-conversion from 795\,nm ($\hat{a}_p$) to 1367\,nm ($\hat{a}_s$)—characterized by $T_p$ and $\eta_s$—we also theoretically explore the up-conversion from 1367\,nm ($\hat{a}_s$) back to 795\,nm ($\hat{a}_p$), described by $T_s$ and $\eta_p$. We find that when the system operates under conditions that optimal conversion efficiency—characterized by a strong coupling field and large coupling detuning— $\eta_p$ and $\eta_s$ are approximately equal~\cite{PoHan}. However, $T_s$ and $T_p$ generally exhibit discrepancies unless the OD is sufficiently large (OD$>$100) and the driving field is strong. This phenomenon can be attributed to the coupling field, which enhances the coherence between states $|1\rangle$ and $|3\rangle$, thereby equalizing the conditions for both up-conversion and down-conversion processes. At lower conversion efficiencies, the V-type and cascade-type EIT mechanisms influence $T_s$ and $T_p$ differently. This duality underscores the versatility of the diamond-type FWM scheme, which not only facilitates the efficient generation of telecom-band photons but also enables their reconversion to the near-infrared regime when required~\cite{CE54}.


\section{CONCLUSIONS}  \label{conclusions}

We employ three incident circularly polarized beams and carefully selected energy levels to establish a cycling transition in a single-Zeeman-sublevel structure. This configuration effectively minimized OD depletion and enabled high conversion efficiency even under lower OD conditions compared to previous studies. We have achieved TFC from 795\,nm to 1367\,nm with conversion efficiencies of 66\% at an OD of 75 and 80\% at an OD of 110 in a diamond-type FWM system using a weak coherent probe field. These results exceed previously demonstrated efficiencies in cold atomic systems using single-photon-level inputs. Furthermore, the measured spectral characteristics of the inherent V-type and cascade-type EIT agree with theoretical predictions, elucidating the underlying mechanisms and indicating that further efficiency improvements can be realized by increasing the OD; for example, at ${\rm OD} = 200$, the conversion efficiency is predicted to approach 90\%~\cite{PoHan}. 

Future work will focus on replacing the weak coherent probe with a true single-photon source~\cite{Shiu1,Shiu2} to directly verify the quantum fidelity of the converted states. To this end, we are actively working to suppress background leakage light and minimize noise from the strong driving field, which could otherwise degrade the quantum-state purity of the frequency-converted photons. While the current implementation operates under weak coherent light conditions, this diamond-type FWM system constitutes a high-efficiency platform with the potential to preserve quantum states, offering a promising pathway toward scalable long-distance quantum communication and advanced quantum information processing.


\section*{ACKNOWLEDGMENTS} \label{acknowledgments}

We sincerely thank Y.-L. Syue, Y.-C. Lee, Y.-L. Tsai, J.-H. He, and S.-Y. Chen for their contributions to the initial setup of the experimental system. This work was supported by the National Science and Technology Council of Taiwan under Grants No. 112-2112-M-006-034 and No. 113-2112-M-006-028. We also acknowledge support from the Center for Quantum Science and Technology within the framework of the Higher Education Sprout Project by the Ministry of Education in Taiwan.




\end{document}